# Measurement of a reaction-diffusion crossover in exciton-exciton recombination inside carbon nanotubes using femtosecond optical absorption

J. Allam[1*], M. T. Sajjad[1,†], R. Sutton[1], K. Litvinenko[1], Z. Wang[1,‡], S. Siddique[1,§], Q-H. Yang[2,#], W. H. Loh[3], and T. Brown[2,¶]

[1]*Advanced Technology Institute, Department of Physics, University of Surrey, Guildford GU2 7XH, UK*

[2]*School of Chemistry, University of Southampton, Highfield, Southampton SO17 1BJ, UK*

[3]*Optoelectronics Research Centre, University of Southampton, Highfield, Southampton SO17 1BJ, UK*

Exciton-exciton recombination in isolated semiconducting single-walled carbon nanotubes was studied using femtosecond transient absorption. Under sufficient excitation to saturate the optical absorption, we observed an abrupt transition between reaction- and diffusion- limited kinetics, arising from reactions between incoherent localized excitons with a finite probability of ~ 0.2 per encounter. This represents the first experimental observation of a crossover between classical and critical kinetics in a 1D coalescing random walk, which is a paradigm for the study of non-equilibrium systems.

PACs: 78.67.Ch, 78.47.+p, 71.35.-y, 05.70.Ln

Many natural processes can be described in terms of diffusive transport of particles which react in proximity, for example chemical and biological reactions, crystallization, epidemics and population growth [1-3]. Diffusion-limited coalescence $A + A \to A$ is a paradigmatic class of reactions, widely studied since Smoluchowski's early work [4]. Despite its apparent simplicity, it exhibits rich kinetic behavior governed by competition between reactive ordering, where coalescing pairs create local fluctuations in density, and diffusive mixing which homogenizes the spatial distribution. In spatial dimensions (*d*) greater than 2, diffusion is efficient and the reaction rate follows classical laws of mass action with mean density *n* decaying as $t^{-1}$ at long times. However diffusion is inefficient for *d* < 2: the diffusive random walk explores space 'compactly' [5], revisiting the same location many times. This leads to anomalously slow decay $n \propto t^{-1/2}$ [6] and to a self-ordered spatial distribution [7] with characteristic scale $\sqrt{Dt}$ where *D* is the diffusion coefficient. The diverging scale at long times is reminiscent of equilibrium critical phenomena, and such reactions are a prototype for the emergence of self-organization and critical behavior in non-equilibrium stochastic systems.

Unusually for non-equilibrium processes, diffusion-limited coalescence is an exactly solvable problem in one spatial dimension (1D) by a variety of theoretical methods [8-11], and is accessible to experimental investigation through exciton-exciton recombination reactions in semiconductors. In exciton-exciton recombination, a pair of singlet excitons (*X*) exchange energy: one is eliminated while the other is excited to a high energy state ($X^*$) before it loses energy to the lattice and returns to the original state: $X + X \to X^* \to X + \text{heat} \to X$. If the exciton cooling and heat dissipation are sufficiently rapid, the kinetics of the recombination process can be approximated by diffusion-limited coalescence. In pioneering work in the late 1980s, Kopelman and coworkers studied the recombination of laser-induced excitons in organic semiconductors where the effective dimensionality was varied in disordered samples [12] and ultrathin wires [13]. Highly anisotropic molecular crystals [14] and more recently semiconducting carbon nanotubes [15] were subsequently investigated. These experiments showed asymptotic diffusion-limited behavior $n \propto t^{-1/2}$ in agreement with theoretical predictions, and represent some of the principal experimental evidence for non-classical fluctuation-dominated kinetics.

In exact theoretical methods the reaction rate is assumed to be infinite, i.e. the reaction probability *p* per encounter is unity. However in real experimental systems, the presence of energy barriers, exclusion mechanisms or orientation dependence act to reduce the reaction probability. In 1985, Kang and Redner [16] used scaling arguments to show that a finite reaction rate gives rise to a regime of reaction-limited behavior at early times and high excitation densities, with a crossover to diffusion-limited kinetics at later times. This was supported by approximate theoretical models [17-19] and Monte Carlo simulations [16,20]. A rigorous justification is found in field theoretic approaches, where the crossover corresponds to a trajectory between different fixed points of the renormalization group scaling transformation [21]. In experiments, a slow crossover might prevent the asymptotic regime from being reached in systems of finite size. In spite of this theoretical interest and experimental relevance, there has been no report (as far as we are aware) of a reaction-diffusion crossover in any experimental realization of a 1D coalescing random walk.

Here, we demonstrate for the first time an unambiguous crossover between reaction- and diffusion- limited scaling regimes, obtained by careful selection of experimental conditions and by the use of carbon nanotubes exhibiting strictly 1D diffusive transport. We illuminate the sample at sufficiently high intensity to saturate fully the optical absorption, avoiding the influence of spatially non-uniform excitation on the initial decay and allowing a study of the intrinsic kinetics at early times. We also show how the saturation condition allows determination of diffusion and reaction parameters without knowledge of the absolute exciton density, which is subject to uncertainty in the absorption cross-section and concentration of the probed species. Semiconducting single walled nanotubes are ideal candidates for the study of 1D exciton recombination [15,22,23]. Nanotube diameters of ~ 1 nm lead to quantum confinement of the transverse motion and transport of excitons is highly one-dimensional [24]. The preparation of isolated nanotubes [25] has allowed extensive studies of exciton photophysics [26] which show that photoexcited excitons are compact, stable against dissociation at room temperature [27] and under high excitation [28], but diffuse with a high mobility [29]. They have sufficiently long radiative lifetimes [30] so that bimolecular interactions can be studied in the absence of single particle processes by using femtosecond pump-probe methods.

We produced an ensemble of isolated HiPco nanotubes [31] wrapped in single-strand DNA and dispersed in water; details of our sample preparation and characteristics are reported in reference [32]. The optical absorption spectrum shows features characteristic of isolated nanotubes and an excitation wavelength of 1133 nm was selected corresponding to nanotube species with diameters 0.8 – 0.9 nm [24]. An average length $L = 184 \pm 5$ nm was measured by atomic force microscopy. A transient population of excitons was generated by illumination with a short light pulse from an amplified Ti:sapphire laser with 250 kHz repetition rate, and the decay was studied by standard degenerate pump-probe techniques. The diameters of focused pump and probe beams were 130 and 95 µm respectively and the sample was contained in a cuvette with 1 mm path length. The energy of the pump pulse was varied between 0.2 and 104 nJ, with the maximum energy corresponding to a fluence of approximately 0.8 mJ cm$^{-2}$. The mean and standard error of multiple scans were determined; weak interference fringes between copolarised pump and probe beams gave rise to a peak in the standard error (inset of Fig. 1(a)), used to identify temporal overlap of the pump and probe beam to within ± 10 fs. The pulse width of 105 fs is a convolution of ≈ 60 fs pump and probe pulses with factors associated with the non-colinear geometry.

Figure 1(a) shows the evolution of the differential transmission $\Delta T / T_0$, which for a given excitation condition is proportional to the total exciton population. The signal amplitudes have been normalized at times > 10ps, showing at long times a decay whose form is independent of excitation density, while at shorter times a rapidly decaying component emerges with increasing excitation. The correct identification of these two regimes is the main purpose of this paper. Figure 1b shows the data on a log-log plot where straight lines correspond to power-law decays. At long times ($t$ > 10 ps), $\Delta T / T_0$ appears to decay with a diffusion-limited $t^{-1/2}$ dependence, but the amplitude varies with excitation in contrast to the expectation for an asymptotic algebraic decay. We attribute this to an increase in the photoexcited volume, illustrating the experimental difficulty in relating $n$ to $\Delta T / T_0$ when the absorption is both nonlinear and non-uniform. The signal amplitude saturates with excitation strength at both long and short times (inset of Fig.

1(b)), as also observed in the early dynamics ($t < 0.1$ ps), where at low excitation the population builds continuously during illumination while at high excitation it saturates at earlier times. The saturation of the optical absorption has been attributed to phase-space filling due to Pauli exclusion of excitons [33].

To quantify the algebraic decay $n \propto t^{\alpha}$, the exponent is determined from $\alpha = (t/n) dn/dt$ and shown in Fig. 2(a). For $t > 10$ ps, a value of -0.5 is obtained for all excitation densities, with an average of $\alpha = -0.51 \pm 0.01$ between 20 and 100 ps, indicating diffusion-limited behavior. For increasing excitation, the magnitude of $\alpha$ at short times increased but did not reach the reaction-limited value of -1; this is a consequence of the constant of integration associated with the initial density $n_0$ at $t = 0$. We show that the early dynamics at high excitation are indeed reaction-limited by examining the underlying rate equation. In Fig. 2(b) we plot $dn/dt$ divided by $n^2$ (the factor relating $n$ to $\Delta T/T_0$ is unimportant here). For the highest excitations (where the absorption is fully saturated), this takes a constant value for times $0.4 < t < 1.2$ ps, confirming the existence of a reaction-limited region governed by classical kinetics. Figure 2 demonstrates the coexistence of reaction- and diffusion- limited behavior occurring at different exciton densities *within the same sample*. Fig. 2(b) also confirms that the reaction is bimolecular, and therefore that the region with $\alpha = -0.5$ arises from 1D diffusion-limited bimolecular kinetics and not from classical kinetics of a free carrier (3-body) Auger process which would have the same kinetic signature.

For further analysis of the data, we employ a simple model based on effective rate equations. In the reaction-limited regime, the classical rate equation $dn/dt = -k_r n^2$ gives a density $n \sim t^{-1}$ at times $t \gg t_1$ where $t_1 = (k_r n_0)^{-1}$ is the timescale for the onset of reactions, $k_r$ is the rate coefficient, and $n_0$ is the initial population. In the diffusion-limited regime, the controlling factor is the time $\sim x^2/4D$ taken for a pair of particles of separation $x$ to meet. Exact solution gives an effective rate equation $dn/dt = -k_d n^3$ valid at times $t \gg t_0$ where $t_0 = (2 k_d n_0^2)^{-1}$ and $k_d = \pi D$, and a density $n$ that decays anomalously slowly as $t^{-1/2}$ [2]. The crossover between these regimes can be approximated by considering each coalescence event as a sequential diffusion and reaction; a similar separation was discussed for diffusion-limited chemical reactions [34]. The time for a pair of particles to coalesce $-n(dt/dn)$ is given by the sum of the time to first encounter $(k_d n^2)^{-1}$ and the additional time to make multiple attempts at reaction $(k_r n)^{-1}$. Integration gives the time $t$ to evolve to density $n$ from the initial density $n_0$

$$t(n) = \frac{1}{2k_d}\left(\frac{1}{n^2}-\frac{1}{n_0^2}\right) + \frac{1}{k_r}\left(\frac{1}{n}-\frac{1}{n_0}\right) . \tag{1}$$

The decay is algebraic once the population has decayed to a value $n \ll n_0$ and there are regimes of reaction- and diffusion-limited behavior at high and low density respectively. Equation (1) can be inverted to give an expression for $n(t)$ equivalent to that obtained from scaling [18] and 'empty-intervals' [19] calculations, which have been shown to agree well with phenomenological models [17] and Monte Carlo simulations [16,20].

Since we do not know the absolute exciton density, we write (1) in terms of a relative density

$$t_s = t + t_0 + t_1 = t_0 (n_0/n)^2 + t_1 (n_0/n) . \tag{2}$$

The constants of integration $t_0$ and $t_1$ have been absorbed into a time shift, yielding a 'scaling time' $t_s$ whose relation to $n$ is independent of the initial condition and is purely algebraic in the high and low density limits. Crossover occurs at $n_2 = (t_0/t_1)n_0$ and $t_2 = t_1^2/t_0$. Figure 3(a) shows a plot of $n/n_0$ against $t_s$ under the highest excitation, where we believe the absorption to be fully saturated giving an initial density $\tilde{n}_0$. In the experiment the delay time is measured from the peak of the excitation pulse, whereas the decay process starts (roughly speaking) from when the excitation is turned *off*; an offset of $0.143 \pm 0.005$ ps was required to reproduce the reaction-limited region as observed in Fig. 2(b). Distinct regions of $t^{-1}$ and $t^{-1/2}$ decay are seen in Fig. 3(a). The characteristic timescales at initial density $\tilde{n}_0$ are $\tilde{t}_0 = 23.9 \pm 0.5$ fs and $\tilde{t}_1 = 303 \pm 3$ fs, and crossover occurs at $t_2 = 3.8 \pm 0.2$ ps. The corresponding decay exponent $\alpha' = (t_s/n)dn/dt_s$ is shown in Fig. 3(b). Between $0.5 \text{ ps} \leq t \leq 100 \text{ ps}$ it exhibits intrinsic behavior with a crossover from $\alpha' = -1$ to $\alpha' = -0.5$. This is compared in Fig. 3(b) to the calculated exponent from (2)

$$\alpha' = -\frac{1 + 2t_s/t_2 + \sqrt{1 + 4t_s/t_2}}{1 + 4t_s/t_2 + \sqrt{1 + 4t_s/t_2}}. \tag{3}$$

The experimental crossover function is significantly more abrupt that that indicated by (3).

Mesoscopic parameters describing diffusion and reaction can be obtained from $\tilde{t}_0$ and $\tilde{t}_1$ by exploiting the saturation condition, in a simple interaction model comprising random walks of particles on a lattice with hopping frequency $t_{hop}^{-1}$, single occupancy on a lattice site, and nearest-neighbor interactions. The lattice spacing represents both exclusion and interaction lengths, and the maximum particle density is therefore the inverse of the lattice spacing ($\tilde{n}_0 = 1/l_{ex}$ where $l_{ex}$ is the axial length of the exciton). The hopping time is $t_{hop} = l_{ex}^2/2D = \pi t_0 (n_0 l_{ex})^2$ which at saturation is $t_{hop} = \pi \tilde{t}_0 = 79 \pm 2$ fs. For finite reaction probability $p$, the number of additional attempts at reaction following the first encounter is $(1/p - 1)$ and the macroscopic reaction coefficient is $k_r = (l_{ex}/t_{hop}) p/(1-p)$. The reaction probability is given by $p/(1-p) = (\pi t_0/t_1)(n_0 l_{ex})$ and at saturation $p = (\pi \tilde{t}_0/\tilde{t}_1)/(1 + \pi \tilde{t}_0/\tilde{t}_1) = 0.19 \pm 0.01$. *On average, a pair of excitons interact ~5 times before recombination occurs*. To confirm the validity of these parameters we determine the exciton dephasing rate associated with thermally-activated hopping $t_{hop}^{-1} = 13 \text{ ps}^{-1}$ and the maximum exciton-exciton dephasing rate $(2/p)/(\tilde{t}_0 + \tilde{t}_1) = 31 \text{ ps}^{-1}$. These are consistent with four-wave mixing experiments on nanotubes of similar dimensions where the rates were $\approx 15 \text{ ps}^{-1}$ and $\approx 35 \text{ ps}^{-1}$ respectively [35].

Observation of crossover requires that the crossover density lies within limits imposed by particle exclusion and finite sample size $l_{ex}^{-1} \gg n_2 \gg L^{-1}$, restricting the reaction probability to $1 \gg p/(1-p) \gg l_{ex}/L$. The quantity $p/(1-p)$ describes the ratio of exciton depopulation to dephasing (i.e. inelastic to elastic scattering). For $p \to 0$, dephasing dominates as for Wannier excitons in inorganic semiconductors, whereas for $p \to 1$ depopulation dominates as for Frenkel excitons in molecular organic materials. This quantifies the observation that the excitonic properties of carbon nanotubes are intermediate [36], leading to dephasing and depopulation rates of similar magnitude [37].

Determination of the macroscopic coefficients $k_r$ and $D$ requires a reliable relation between $\Delta T/T_0$ and $n$. In its absence, upper bounds can be obtained by estimating the maximum exciton density. The $t^{-1/2}$ decay associated with exciton-exciton recombination persists for times up to at least 100 ps, so that there remain at least 2 excitons per nanotube. The ratio of $\Delta T/T_0$ at 100 ps and 0.1 ps yields an initial population $N_{max} \geq 120$, and hence the exciton length is $l_{ex} = L/N_{max} \leq 1.5$ nm, similar to theoretical values [27]. The diffusion coefficient is $D = l_{ex}^2/2t_{hop} \leq 0.15$ cm$^2$s$^{-1}$, indicating a diffusive environment comparable to earlier reports [29,33], i.e. exciton transport within the present samples is not exceptional. However, our reaction coefficient $k_r = l_{ex}/t_1 \leq 5.0$ nm ps$^{-1}$ is smaller than an earlier estimate by a factor > 60 [22].

In earlier studies of exciton-exciton recombination on nanotube samples similar to those reported here, an initial $t^{-1}$ decay under high excitation was reported. This was followed by a slower decay which was attributed to exponential *single exciton* decay processes [22,23] rather than to diffusion-limited exciton-exciton recombination, leading to an assumption of low exciton densities. In order to explain the fast initial decay, Wang et al. proposed a giant enhancement of the exciton-exciton Coulomb interaction by 1D confinement [22]. Here we reach the opposite conclusion: it is a *reduction* in reaction probability ($p < 1$) that leads to reaction-*limited* classical kinetics. Ma et al. attributed the $t^{-1}$ decay to coherent interactions between extensively delocalized excitons [23]. However we have determined the exciton length (< 1.5 nm) and exciton coherence time (< 80 fs) to be short, justifying the use of a classical stochastic description of these quantum-confined excitations. On the other hand, long-range electrostatic interactions are not ruled out; indeed Förster resonant energy transfer through intra-tube dipolar interactions seems the most likely interaction mechanism [38].

Existing theories and numerical simulations of reaction-diffusion systems are based on highly simplified models of particle interactions and transport. The abrupt crossover observed in our experiment suggests that the models may be insufficient away from the asymptotic limit; this is a topic for further research. Our results identify exciton reactions on carbon nanotubes as an experimental platform permitting precise investigations of anomalous reaction kinetics in a low-dimensional non-equilibrium stochastic system.

We thank C. Giusca for performing the AFM measurements and J. D. Carey for helpful discussions. The work was supported by the UK Engineering and Physical Sciences Research Council 'Next Generation Electrophotonics' programme (grants EP/C010531/1 and EP/C010558/1).

*Corresponding author, *j.allam@surrey.ac.uk*

[†]Present address: Organic Semiconductor Centre, University of St Andrews, North Haugh, St. Andrews, KY169SS, UK.

[‡]Present address: Research Centre of Quantum Engineering and Technology, Shanghai Advanced Research Institute, CAS, Shanghai 201203, China.

[§]On leave from: Department of Physics, University of Engineering and Technology, Lahore, 54890, Pakistan

[#]Present address: Key Laboratory of Thermal Management Engineering and Materials, Graduate School at Shenzhen, Tsinghua University, Shenzhen, 518055, China.

[¶]Present address: Department of Chemistry, University of Oxford, Oxford, OX1 3TA, UK

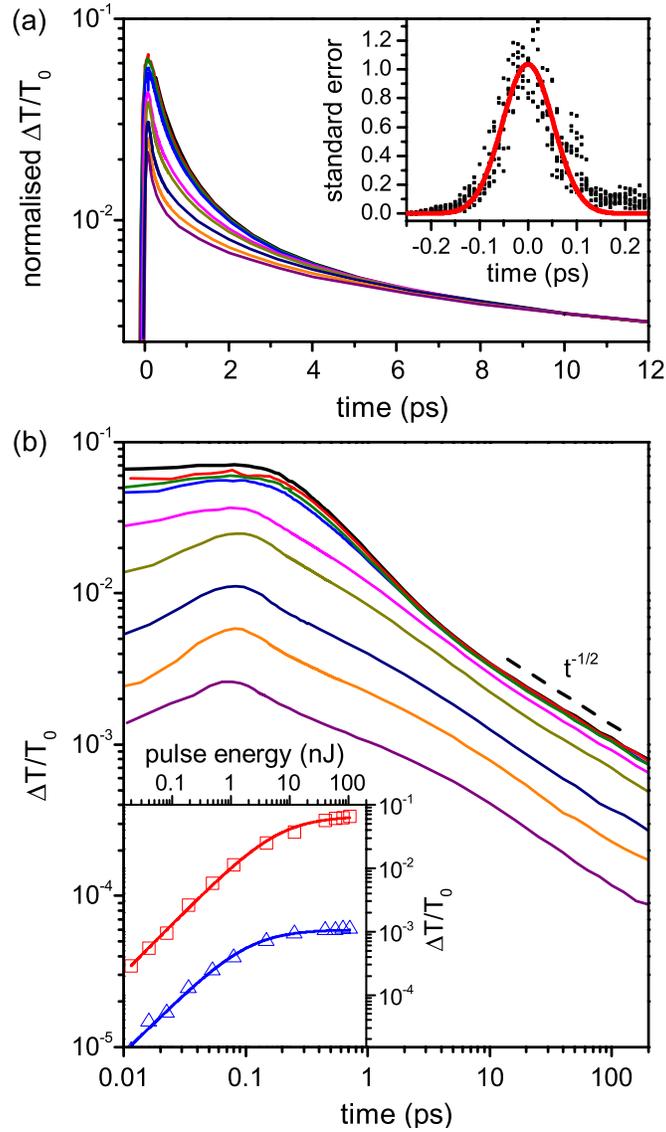

FIG. 1. (Color online) (a) Decay of transmission $\Delta T / T_0$, normalized at times > 10 ps, at pulse energies (bottom to top) of 0.19, 0.48, 1.1, 4.0, 12, 40, 60, 80, 104 nJ. The inset shows the standard error from multiple scans (symbols), and the line is a Gaussian fit of width 105 fs. (b) The same data (not normalized) on a log-log plot. The dashed line shows the asymptotic $t^{-1/2}$ decay. The inset shows the differential transmission at delays of 0.1 ps (squares) and 100 ps (triangles) as a function of pulse energy.

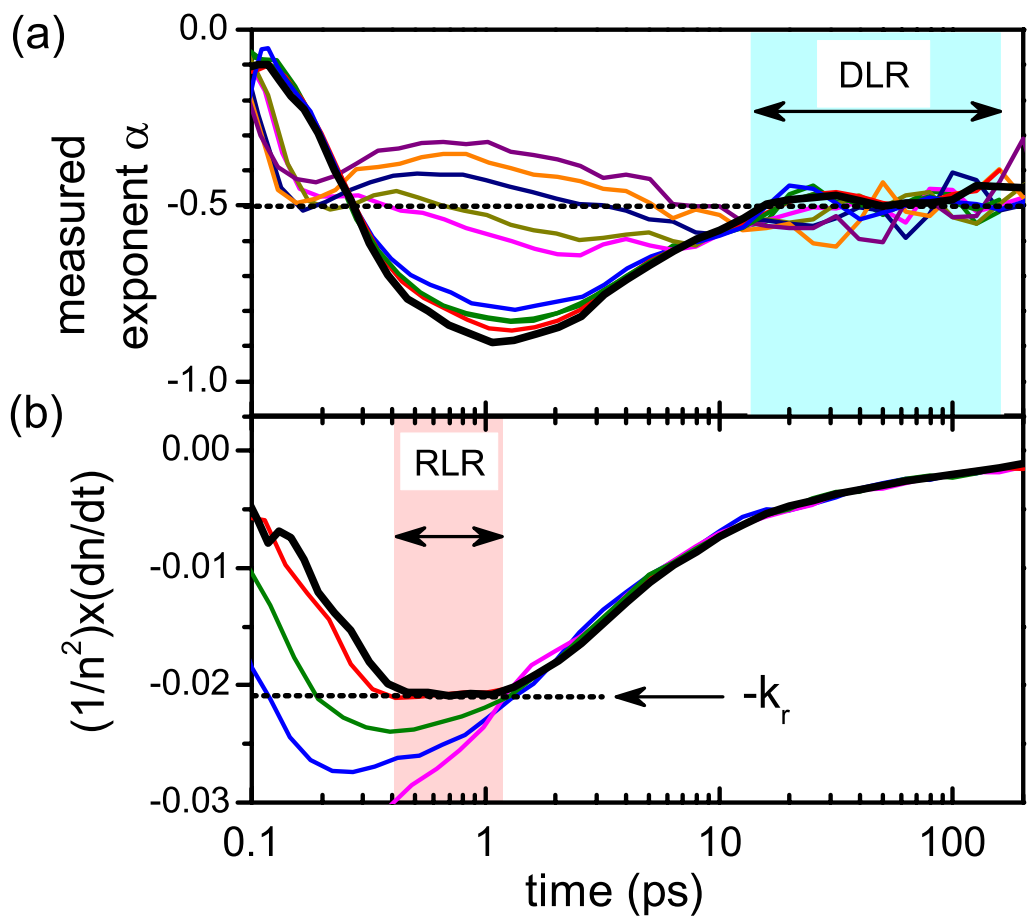

FIG. 2. (Color online) (a) The decay exponent $\alpha$ at identical pulse energies as Fig. 1 (top to bottom). $\alpha = -0.5$ indicates the diffusion-limited regime (DLR). (b) The reaction rate divided by $n^2$ (in arbitrary units, 104 nJ pump pulses) showing a constant value for $0.4 < t < 1.2$ ps, indicating a reaction-limited regime (RLR).

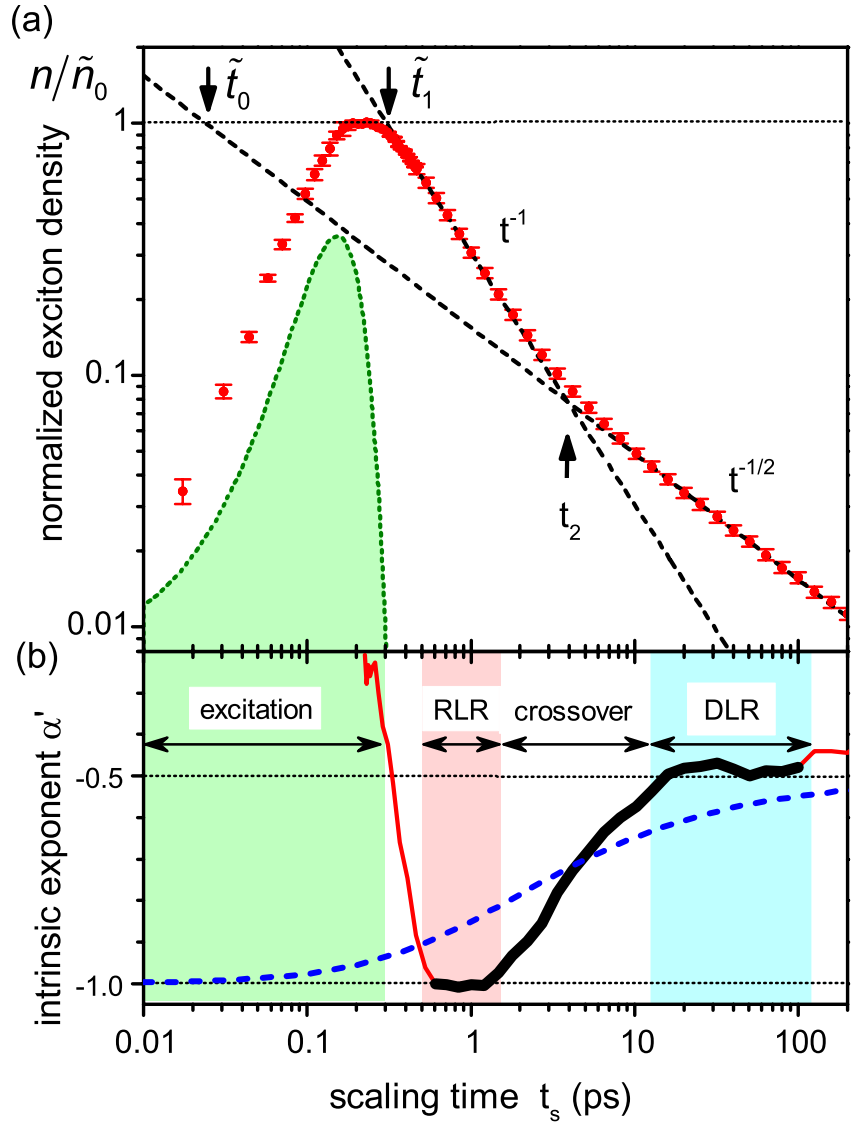

FIG. 3. (Color online) (a) Normalised exciton density (symbols) for 104 nJ pump pulses, plotted against the scaling time $t_s$. The error bars represent the standard error. The dotted curve shows the pump pulse (the Gaussian fit from Fig. 1(a) with arbitrary amplitude). The dashed lines indicate $t^{-1}$ and $t^{-1/2}$ decay, and the arrows indicate the characteristic times $\tilde{t}_0$, $\tilde{t}_1$ and $t_2$. (b) Decay exponent for the data in Fig. 3(a). The heavy line shows the experimentally-determined reaction-diffusion crossover function, and the dashed line is the prediction of (3).